\documentclass[conference]{IEEEtran}
\IEEEoverridecommandlockouts
\usepackage{cite}
\usepackage{amsmath,amssymb,amsfonts}
\usepackage{algorithmic}
\usepackage{graphicx}
\usepackage{textcomp}
\usepackage{xcolor}
\def\BibTeX{{\rm B\kern-.05em{\sc i\kern-.025em b}\kern-.08em
    T\kern-.1667em\lower.7ex\hbox{E}\kern-.125emX}}

\usepackage[compatibility=false]{caption}
\DeclareCaptionFont{quackfont}{\fontfamily{ptm}\fontsize{7.5pt}{9pt}\selectfont}
\usepackage[labelformat=simple,font=quackfont]{subcaption}

\DeclareUnicodeCharacter{2212}{\ensuremath{-}}

\makeatother

\begin{document}

\title{Deep SIMO Auto-Encoder and Radio Frequency Hardware Impairments Modeling for Physical Layer Security\\
\thanks{This work was supported by the 5G Positioning, Sensing and Security Functions (5G-PSS) through Business Finland under Grant 6868/31/2021 and also is part of the DIOR project that has received funding from the European Union’s MSCA RISE programme under grant agreement No. 10100828.}
}
\author{Abdullahi~Mohammad,~\IEEEmembership{Member,~IEEE,}
   Mahmoud Tukur Kabir,~\IEEEmembership{Member,~IEEE,}\\
    Mikko~Valkama,~\IEEEmembership{Fellow,~IEEE} and 
    Bo~Tan,~\IEEEmembership{Member,~IEEE}\\
     \IEEEauthorblockA{\IEEEauthorrefmark{1}Tampere Wireless Research Centre, Faculty of Information Technology and Communication Sciences,\\
     Tampere University, Finland}\\[-2.0ex]
e-mail: (abdullahi.mohammad; mahmoud.kabir; mikko.valkama; bo.tan)@tuni.fi
}

\maketitle

\begin{abstract}
This paper presents a novel approach to achieving secure wireless communication by leveraging the inherent characteristics of wireless channels through end-to-end learning using a single-input-multiple-output (SIMO) autoencoder (AE). To ensure a more realistic signal transmission, we derive the signal model that captures all radio frequency (RF) hardware impairments to provide reliable and secure communication. Performance evaluations against traditional linear decoders, such as zero-forcing (ZR) and linear minimum mean square error (LMMSE), and the optimal nonlinear decoder, maximum likelihood (ML), demonstrate that the AE-based SIMO model exhibits superior bit error rate (BER) performance, but with a substantial gap even in the presence of RF hardware impairments. Additionally, the proposed model offers enhanced security features, preventing potential eavesdroppers from intercepting transmitted information and leveraging RF impairments for augmented physical layer security and device identification. These findings underscore the efficacy of the proposed end-to-end learning approach in achieving secure and robust wireless communication.
\end{abstract}


\section{Introduction}
\IEEEPARstart{T}{he} demand for massive wireless device connectivity and reduced end-to-end latency requirements are catalysts for deploying 6G technologies. As a result of this massive device connectivity, security concerns have garnered significant attention in wireless communication systems due to wireless channels' inherent broadcasting nature. However, the assurance of security in traditional wireless communication systems typically hinges on higher-level protocols' encryption and authentication mechanisms, which are inevitably characterized by high computational complexity \cite{edman2016security}. Numerous studies have been conducted on physical layer (PHY) security solutions to explore alternative, more energy-efficient security solutions. These solutions primarily leverage the inherent characteristics of wireless propagation channels or transceivers \cite{moualeu2019physical,sun2020end,illi2023physical}. In recent years, there has been a surge in interest in end-to-end learning communication, revealing the promise of energy-efficient and secure wireless communication paradigms \cite{o2017introduction,mohammad2023learning}. Classical communication theory traditionally decomposes systems into modules, each handling a specific signal processing. Conversely, end-to-end learning communication, based on deep learning (DL) principles, dispenses this modular approach and aims to discern optimal communication strategies within particular wireless environments autonomously. By jointly generating encoding and decoding schemes at the sender and receiver, respectively, this approach ensures that symbols generated by the sender can only be perfectly deciphered by the intended recipient. Furthermore, owing to the intricate nature of neural network (NN) models, their parameter space significantly surpasses optimal problem solutions. With challenges in parameter initialization, learning rate calibration, and the array of training methodologies, achieving complete consistency across NN models and outputs becomes a formidable task. Consequently, end-to-end learned communication systems inherently exhibit potential characteristics of natural endogenous security.

DL is used to design and optimize beamformers for  multiple-input multiple-output (MIMO) and SIMO wiretap channels, demonstrating superior robustness and secure capacity compared to analytical and numerical approaches \cite{zhang2019deep,yun2020deep}. However, the designs lack a more realistic RF signal modeling that could present a comprehensive framework for secure transmission scenarios. For instance, reference \cite{fritschek2019deep} proposes an innovative approach to confidential message transmission over Gaussian wiretap channels by introducing a secure loss function based on cross-entropy. Despite its promise, research efforts in secure communication via end-to-end learning using AE designs for MIMO and SIMO wiretap channels remain limited. In this correspondence, we propose a SIMO deep AE design considering all the RF impairments at the RF transmission chains to present a more realistic signal transmission. Our contribution as summarized as follows:
\begin{itemize}
    \item Departing from conventional methodologies that treat RF hardware impairments merely as additive random components, we present a derived and holistic RF signal model encompassing all pertinent hardware impairments at the transmitter.
    \item We introduce an AE-based framework for SIMO wiretap channels. Our proposed model is trained using a meticulously designed joint weighted loss function aimed at fortifying resilience against eavesdropping attempts while concurrently maintaining the fidelity of communication to the legitimate receiver. 
\end{itemize}
\section{RF Hardware Impairments Signal Model}
The transmitted signal is susceptible to various distortions throughout the transmission chain due to imperfections in the physical RF hardware components. At the transmitter, the digital base-band signal, denoted as $x[n]$, is converted to a continuous analog baseband signal $x(t)$ through the digital-to-analog converter (DAC). The signal $x(t)$ then undergoes a series of signal processing steps involving up-conversion and amplification using RF circuits, including oscillators and power amplifiers (PA). These components exhibit distinct impairments, and their collective influence is embedded in the RF signal. 

\subsection{DAC Non-linearities}
The DAC is affected by finite digital input precision, which can introduce nonlinearities that may vary across its units. Consequently, the transmitter can be characterized by a set of parameters that uniquely describe its components' input/output characteristics. The nonlinearity induced by the DAC can be modeled using a Taylor series expansion around 0 up to a maximum degree of $k_{max}$ \cite{balatsoukas2015baseband}. The DAC accepts the in-phase ($x_{I}[n]$) and quadrature of the modulated digital signal ($x_{Q}[n]$) and outputs their corresponding analog versions, expressed as
\begin{equation}
    x_{I}(t)=\sum_{k=1}^{k_{max}}\rho_{1, k}\Re \left \{ x[n] \right \}^{k}
\end{equation}
\begin{equation}
 x_{Q}(t)=\sum_{k=1}^{k_{max}}\rho_{2, k}\Im \left \{ x[n] \right \}^{k}
\end{equation}
where $\rho_{i, k}
\in\ \mathbb{R}; \forall i\ \{1, 2\}$ describes the DAC's induced non-linearity; $\ k\in \{ 1,\cdots,k_{max}\}$. Therefore, the continuous time complex base-band $x(t)$ signal is obtained form the in-phase (I) and quadrature (Q) branch as
\begin{equation}
        x(t)=\sum_{k=1}^{k_{max}}\rho_{1, k}\Re \left \{ x[n] \right \}^{k}+j\sum_{k=1}^{k_{max}}\rho_{2,k}\Im \left \{ x[n] \right \}^{k}
\end{equation}
For the sake of simplicity, we assume that the DACs are perfectly matched, resulting in equal weighting coefficients for the I/Q branches ($\rho_{1,k}=\rho_{2,k}=\rho_{k}$) \cite{balatsoukas2015baseband}, simplifying the complex analog base-band signal representation. Thus, the complex analog base-band signal becomes:
\begin{equation}
    x(t)=\sum_{k=1}^{k_{max}}\rho_{k}\left ( \Re \left \{ x[n] \right \}^{k}+j \Im \left \{ x[n] \right \}^{k}\right )
\end{equation}

\subsection{Oscillator and Mixer Imperfections}
The analog signal from the DAC undergoes two up-conversions before being transformed into a bandpass signal as shown in Fig \ref{fig:system_model}. In the initial conversion stage, the signal passes through a mixer and an image rejection filter for intermediate frequency translation, enhancing frequency selectivity. During the second up-conversion, the signal is converted to a bandpass via a secondary mixer, restricting the output signal's bandwidth to the minimum required for transmitting it at the desired carrier frequency. 

\textbf{1. First-Stage Up-conversion:} Imperfections in the LO primarily arise from carrier frequency offset (CFO) $\delta f_{c}$ and phase noise (PN) $\psi(t)$, both essential in modeling the LO's distortions. CFO represents the actual oscillator frequency deviation from the nominal (reference) frequency. Given the nominal frequency $f_{c}^{0}$, the carrier frequency can be expressed as
\begin{equation}
    f_{c}=f_{c}^{0}+\delta f_{c}
\end{equation}
It has been shown in \cite{zhang2021radio} that $\delta f_{c}$ a function of frequency stability ($f_{ppm}$), typically expressed in parts per million (ppm) as:
\begin{equation}
    -\frac{f_{ppm}}{10^{6}}\times f_{c}^{0}\leq \delta f_{c}\leq +\frac{f_{ppm}}{10^{6}}\times f_{c}^{0}
\end{equation}
Without loss of generality, $\psi (t)$ can be added to the instantaneous phase of the signal as follows:
\begin{equation}
    x_{LO}(t)=\cos (\omega_{c}t+\psi (t))
\end{equation}
where $\omega_{c}=2 \pi (f_{c}^{0}+\delta f_{c})$.

The imbalance LO signals at the I and Q arms used for up-conversion are given by 
\begin{equation}
    b_{I}(t)=\cos (\omega_{c}t+\theta)
\end{equation}
\begin{equation}
    b_{Q}(t)=g_{T}\sin (\omega_{c}t-\theta)
\end{equation}
where $g_{T}$ and $\theta$ are gain and phase mismatch or error and follow random uniform distribution in the range of [-1, 1] dB and [-5 5] degrees, respectively. The RF baseband signal in the presence of I/Q imbalance and oscillator imperfections can be written as follows \cite{zhu2013blind}
\begin{equation}
    \text{x}_{RF}(t)={x}_{I}(t)\cos(\omega_{c}t+\psi (t)+\theta)-g_{T}{x}_{Q}(t)\sin(\omega_{c}t+\psi (t)-\theta)
\end{equation}
\begin{multline}\label{x_RF1}
    \text{x}_{RF}(t)=e^{j(\omega_{c}t+\psi(t))}\left [ {x}_{I}(t)+jg_{T}{e}^{j\theta}{x}_{Q}(t) \right ]+\\
    e^{-j(\omega_{c}t+\psi(t))}\left [ {x}_{I}(t)-jg_{T}{e}^{-j\theta}{x}_{Q}(t) \right ]
\end{multline}
Rearranging (\ref{x_RF1}), we have
\begin{multline}\label{x_RF2}
     \text{x}_{RF}(t)=\left [x_{I}(t)\cos \theta +g_{T} x_{Q}(t)\sin \theta \right ]\cos (\omega _{c}t+\psi(t))-\\
\left[x_{I}(t)\sin \theta +g_{T} x_{Q}(t)\cos \theta \right ]\sin (\omega _{c}t+\psi(t))
\end{multline}
By introducing $u_{I}(t)$ and $u_{Q}(t)$ notations for the baseband signal at the I/Q branches, (\ref{x_RF2}) becomes:
\begin{equation}\label{x_rf_real}
    \text{x}_{RF}(t)=u_{I}(t)\cos (\omega _{c}t+\psi(t))-u_{Q}(t)\sin (\omega _{c}t+\psi(t))
\end{equation}
where 
\begin{equation}\label{u_I}
    u_{I}(t)=x_{I}(t)\cos \theta +g_{T} x_{Q}(t)\sin \theta 
\end{equation}
\begin{equation}\label{u_Q}
    u_{Q}(t)=x_{I}(t)\sin \theta +g_{T} x_{Q}(t)\cos \theta
\end{equation}
Following this, it is easy to obtain the equivalent baseband signal as: 
\begin{equation}\label{u_BB}
    u_{BB}(t)=u_{I}(t)+ju_{Q}(t)
\end{equation}
Using (\ref{u_I}) and (\ref{u_Q}), (\ref{u_BB}) can be written as:
\begin{equation}
    u_{BB}(t)=x_{I}(t)e^{j\theta}+jg_{T}x_{Q}(t)e^{-j\theta }
\end{equation}
The resultant RF signal from the mixer and LO is thus 
\begin{equation}
    {\text{x}}_{RF}=\left [ \cos (\omega_{c}t+\psi (t)) +j\sin (\omega_{c}t+\psi (t)) \right ]u_{BB}(t)
\end{equation}
\begin{equation}\label{x_rf_sinal}
    {\text{x}}_{RF}=u_{BB}(t)e^{j(\omega_{c}t+\psi(t)}
\end{equation}
It can be easily seen that (\ref{x_rf_real}) is the real part of (\ref{x_rf_sinal}).

\textbf{2. Second-Stage Up-conversion:} In the second up-conversion, voltage control oscillator (VCO) is used to further up-convert the signal and translate it into bandpass RF signal using bandpass filter. By following similar steps as in the first up-conversion, the output of the second mixer and the VCO is given by
\begin{equation}
{\text{x}}_{BP}(t)=\Re \left \{ {\text{x}}_{RF}(t) \right \}e^{j\theta_\text{vco}}+jg_\text{vco}\Im \left \{ {\text{x}}_{RF}(t) \right \}e^{-j\theta_\text{vco}}
\end{equation}

\begin{equation}
\bar{\text{x}}_{RF}(t)={\text{x}}_{BP}(t)e^{j(\omega _{\text{vco}}t+\psi_{\text{vco}}(t))}
\end{equation}
\begin{equation}
\omega_{\text{vco}}=2\pi \left ( K_\text{vco} \cdot V_\text{vco}+f_\text{vco}^{0} \right )
\end{equation}
where $g_{\text{vco}(t)}$ and $\psi_{\text{vco}(t)}$ are the VCO's gain and phase noise, respectively. $V_\text{vco}$ is the control voltage, whose value depends on the type of VCO, $K_\text{vco}$ is frequency sensitivity gain and $f_\text{vco}^{0}$ is the oscillator's reference or nominal frequency.
Therefore, the amplitude of the transmitted RF bandpass signal is thus
\begin{equation}
    \left | \bar{\text{x}}_{RF}(t) \right |=\left | {\text{x}}_{BP}(t) \right |
\end{equation}
So, the resultant RF bandpass signal $\bar{\mathrm{x}}_{RF}(t)$ can be represented as:
\begin{equation}
    \bar{\mathrm{x}}_{RF}(t)=\left |\bar{\mathrm{x}}_{RF}(t)  \right |e^{j(\omega _{\text{vco}}t+\psi_{\text{vco}}(t))}
\end{equation}

\subsection{Power Amplifier Imperfection}
The PA is an essential transmitter component that boosts low-power signal to a higher amplitude, yet its inherent non-linearity introduces signal distortion. It is characterized by both amplitude-to-amplitude (AM-AM) and amplitude-to-phase (AM-PM) distortions. The widely employed non-linear model for characterizing the non-linear behaviors of PA in a narrowband system is the Saleh model\cite{saleh1981frequency}. This model is expressed in terms of its AM-AM and AM-PM characteristics, described as:
\begin{equation}\label{s_model}
    \text{x}_{PA}(t)=A(\left | {\text{x}}_{BP}(t) \right |)e^{j(\varphi(t) +\Phi(\left | {\text{x}}_{BP}(t) \right |))}
\end{equation}
were $A\left ( \cdot \right )$ and $ \Phi \left ( \cdot \right )$ represent AM/AM and AM/PM effects, respectively, and $\varphi(t) = \angle {\text{x}_{BP}(t)}+\omega _\text{vco}t+\psi_\text{vco}(t)$. 
Without loss of generality, the AM-AM and AM-PM characteristics of Saleh model are respectively defined as follows:
\begin{equation}\label{s_model_amplitude}
    A(\left | {\text{x}}_{BP}(t) \right |)=\frac{\alpha _{A}\left | {\text{x}}_{BP}(t) \right |}{1+\beta _{A}\left | {\text{x}}_{BP}(t) \right |^{2}},
\end{equation}
\begin{equation}\label{s_model_phase}
    \Phi(\left | {\text{x}}_{BP}(t) \right |)=\frac{\alpha _{P}\left | {\text{x}}_{BP}(t) \right |^{2}}{1+\beta _{P}\left | {\text{x}}_{BP}(t) \right |^{2}}
\end{equation}
The $\alpha_{A}$ and $\alpha_{P}$ in (\ref{s_model_amplitude}) and (\ref{s_model_phase}) are amplitude and phase gain factors, respectively, and $\beta _{A}$ and $\beta _{P}$ are the amplitude and phase compression factors, respectively. These factors are the ﬁtting parameters for the measured PA’s AM–AM
characteristics $A(\left | {\text{x}}_{BP}(t) \right |)$ and AM–PM characteristics $\Phi(\left | {\text{x}}_{BP}(t) \right |)$ \cite{schreurs2008rf}. After passing through sequence of RF hardware chains, the transmitted signal is ready to be emitted by the antenna. For simplicity, we drop the subscript and (\ref{s_model}) can be written as:
\begin{equation}\label{s_model_1}
    \text{x}(t)=\bar{\text{x}}(t)e^{j(\omega _\text{vco}t+\psi_\text{vco}(t))}
\end{equation}
where $\bar{\text{x}}(t)=A(\left | {\text{x}}_{BP}(t) \right |)e^{j(\angle {\text{x}_{BP}(t)} +\Phi(\left | {\text{x}}_{BP}(t) \right |))}$.

\begin{figure}[!tp]
     \centering
     \begin{subfigure}[t]{0.241\textwidth}
         \centering
         \includegraphics[width=\linewidth]{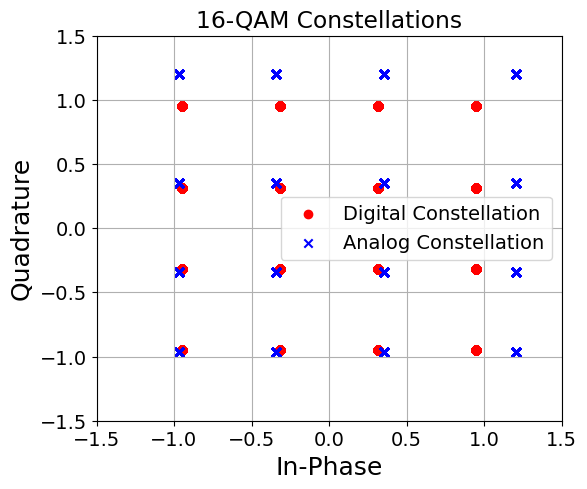}
         \caption{}
     \end{subfigure}
     \hfill
     \begin{subfigure}[t]{0.241\textwidth}
         \centering
         \includegraphics[width=\linewidth]{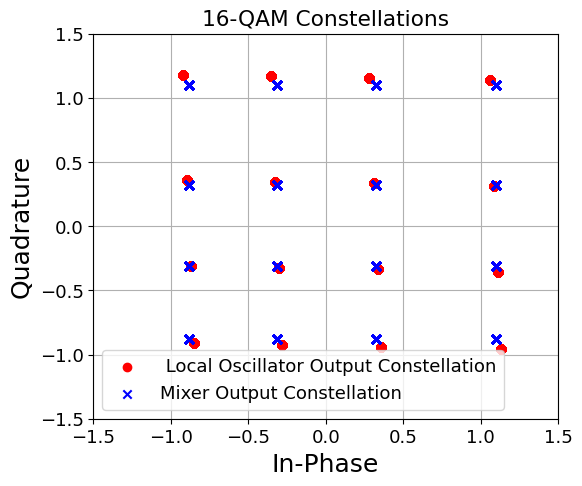}
         \caption{}
     \end{subfigure}
     \hfill
    \begin{subfigure}[t]{0.241\textwidth}
        \centering
        \includegraphics[width=\linewidth]{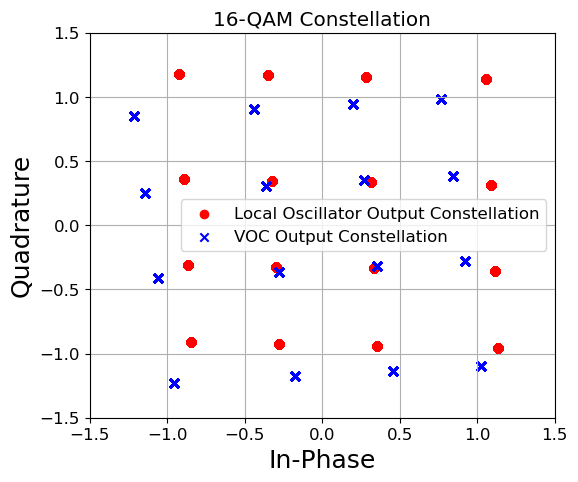}
        \caption{}
    \end{subfigure}
    \hfill
    \begin{subfigure}[t]{0.241\textwidth}
        \centering
        \includegraphics[width=\linewidth]{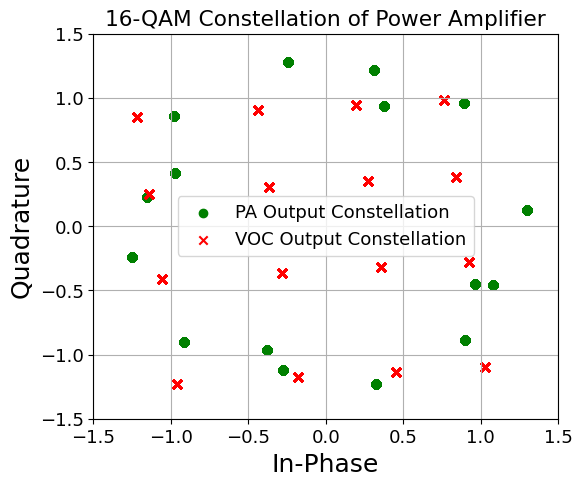}
        \caption{}
    \end{subfigure}
    \caption{(a). Constellation change due to DAC non-linearity; (b). First stage up-conversion symbol constellation due to local oscillator; (c). Second stage up-conversion symbol constellation due to VCO; (d). RF-band symbol constellation due to PA.}
    \label{fig:constellation}
\end{figure}

One of the most crucial parameters influencing the nonlinear behavior of DAC is Integral Nonlinearity (INL). INL defines the deviation between the actual output level of the DAC and its ideal output level given a specific input. For simplicity, this behavior is captured by nonlinear coefficients ($\rho$), as described in Section A. The impact of the DAC's nonlinearity is depicted in Fig. \ref{fig:constellation} (a). As illustrated, the misalignment of the symbol constellation from the DAC with its input digital constellation arises from the nonlinear characteristics of the DAC, whose severity is determined by the degree of the nonlinear coefficients. Similarly, Fig. \ref{fig:constellation} (b) compares the impact of local oscillator's imperfection due to the IQ imbalance in mixer and the phase Similarly, Fig. \ref{fig:constellation} (b) shows the impact of local oscillator's imperfection due to the IQ imbalance in mixer, the CFO and phase noise and the output from the DAC. It can be observed that these imperfections result in constellation rotation. The severity of the RF distortion is also observed in the second up-conversion due to additional nonlinearity introduced by the VCO as shown in Fig. \ref{fig:constellation} (c). Finally, the impact of PA non-linearity is shown in Fig. \ref{fig:constellation} (d), depicting its severity on the symbol constellation.

\section{Physical Layer System Model}
\begin{figure*}[ht]
    \centering
    \includegraphics[width=6.8in,height=1.51in]{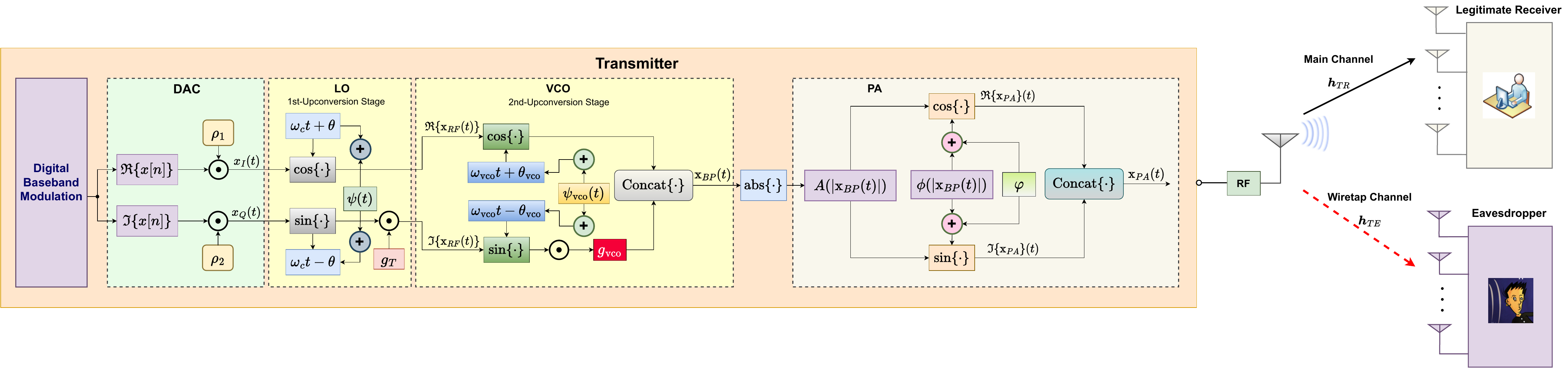}
    \caption{SIMO Transmitter RF chain with Hardware Impairments}
    \label{fig:system_model}
\end{figure*}

Consider a communication system comprising a transmitting node, $T$, which aims to transmit a signal to the intended receiver, $R$, while contending with the presence of an eavesdropper, $E$. The eavesdropper attempts to decode the transmitted signal by intercepting the broadcast from node $T$ to node $R$. Suppose the transmitter, intended receiver, and eavesdropper are equipped with uniform linear arrays of $N_{t}$, $N_{r}$, and $N_{e}$ antennas, respectively. The multipath channel between the different nodes can be geometrically represented as follows:
\begin{equation}            \boldsymbol{h}_{AB}=\sum_{l=1}^{L^{[AB]}}\hbar_{l}\boldsymbol{g}_{r}^{[B]}(\phi_{r,l}){g}_{t}^{[A]}(\phi_{t,l})
\end{equation}
where $ A=T$ and $ B=R$ or $E$ such that $\boldsymbol{h}_{TR}$ and $\boldsymbol{h}_{TE}$ are the channels between the transmitter-intended receiver and transmitter-eavesdropper, respectively as depicted in Fig. \ref{fig:system_model}. Here, $L^{[AB]}$  denotes the number of multipath channels between $A$ and $B$, $\hbar_{l}\sim \mathbb{C}\mathcal{N}(0,1)$ is the complex channel gain, ${g}_{t}^{[A]}\in \mathbb{C}$ and $\boldsymbol{g}_{r}^{[B]}\in \mathbb{C}^{N_{r} \times 1}$ are the steering vectors of the transmitter and receiver, respectively, and $\phi_{r}$ is the angle of arrival (AoA), which follow uniform distribution $\sim \mathcal{U}(-\frac{\pi}{2},\frac{\pi}{2})$, The steering vector is given by:{}
\begin{equation}
\boldsymbol{g}_{r}^{[B]}(\phi)= \frac{1}{\sqrt{N}}\left [
1, e^{-j2 \pi\frac{\bar{d}_{c}}{\lambda_{c}}\sin(\phi_{r})},\cdots,e^{-j2 \pi (N-1)\frac{\bar{d}_{c}}{\lambda_{c}}\sin(\phi_{r})} \right ]^{T}
\end{equation}
where $N\in \left \{ N_{r},N_{e}\right \} $, $\phi\in \left \{\phi_{r}\right \}$, $\bar{d}_{c}$ and $\lambda_{c}$ are the antenna spacing and carrier wavelength. 

For simplicity, we assume that the legitimate receiver and eavesdropper are equipped with $\bar{N}$ antennas, such that $\bar{N}=N_{r}=N_{e}$. The transmit signal, $\text{x}$ will be used instead of \text{x}$(t)$ for the rest of the paper. The received signal vectors at the legitimate receiver and eavesdropper are respectively as given as
\begin{equation}\label{legit_receiver}
    \textbf{y}_{r}=\mathbf{h}_{TR} \ \text{x}+\mathbf{n}_{r}
\end{equation}
\begin{equation}\label{eves_receiver}
    \textbf{y}_{e}=\mathbf{h}_{TE} \ \text{x}+\mathbf{n}_{e}
\end{equation}
where $\text{x} \in \mathbb{C}$ is the impaired transmitted symbol, $\mathbf{h}_{TR}\in \mathbb{C}^{\bar{N}\times 1}$ and $\mathbf{h}_{TE}\in \mathbb{C}^{\bar{N}\times 1}$ refer
to the channel vectors from transmitter to legitimate receiver and eavesdropper, respectively; $\mathbf{n}_{r}\sim \mathcal{N}(0,\sigma_{r}^{2}\mathbf{I}) \in \mathbb{C}^{\bar{N}_{r}}$ and $\mathbf{n}_{e}\sim \mathcal{N}(0,\sigma_{e}^{2}\mathbf{I}) \in \mathbb{C}^{\bar{N}_{r}}$ are the received additive Gaussian
noise vectors at the legitimate receiver and eavesdropper, respectively.

\section{Autoencoder-based SIMO Architecture}
An autoencoder (AE) end-to-end learning for a single-input single-output system (SISO) was first proposed in \cite{o2017introduction}. End-to-end learning entails implementing a transceiver using a pair of multi-layer neural networks (NNs), denoted as $\mathcal{F}_{\Theta_{T}}:\mathcal{M} \to\mathbb{C}$ and $\mathcal{F}_{\Theta_{R}}:\mathbb{C} \to\left [ 0,1 \right ]^{\mathcal{M}}$, which encode and decode the transmitted and received signals (messages), respectively. Here, $\mathcal{F}_{\Theta_{T}}$ and $\mathcal{F}_{\Theta_{R}}$ are the transmitter and receiver parameters. The same principle can be extended to MIMO encoding and decoding for signal detection \cite{o2017physical,song2020benchmarking}. Accordingly, the AE-based SIMO implementation is formulated as follows
\begin{enumerate}
    \item \textbf{Transmitter:} 
    $\mathcal{F}_{\Theta_{T}}:\mathcal{M}^{K}  \to\mathbb{C}^{K}$ maps $K$ consecutive $\mathcal{M}$ constellation points (messages), $\text{x}=\left \{ \text{x}_{1},\cdots,\text{x}_{K} \right \}^{K}\in \mathcal{M}^{K}$ into $K$  coded vector with an average power constraint $\mathbb{E}\left \{ \text{x}{\text{x}}^{\ast} \right \}\le P_{T}$, where $P_{T}$ denotes the total transmission power.
    \item \textbf{Receiver:} $\mathcal{F}_{\Theta_{R}}: \mathbb{C}^{K \times N_{r}} \times \mathbb{C}^{N_{r}} \to \left [ \textsl{X} \right ] ^{\mathcal{M}^{K}}$. The transmitted signal is recovered from the received signal by mapping the learned constellation using the following transformation:
    \begin{subequations}
        \begin{equation}
        \label{eq-a}
        \mathbf{y}=\mathcal{F}_{\Theta_{R}}\left (\mathbf{y} , \mathbf{h}\right ),
        \end{equation}
         \begin{equation}
        \label{eq-b}
        \hat{\text{x}}=\underset{\text{x}\in \mathcal{M}^{K}}{\text{argmax}\ \left \{ \mathbf{y} \right \}}
        \end{equation}
    \end{subequations}
    where $\hat{\text{x}}$ is the recovered symbol.
\end{enumerate}
\begin{figure}[ht]
    \centering
    \includegraphics[width=3.4in,height=2.4in]{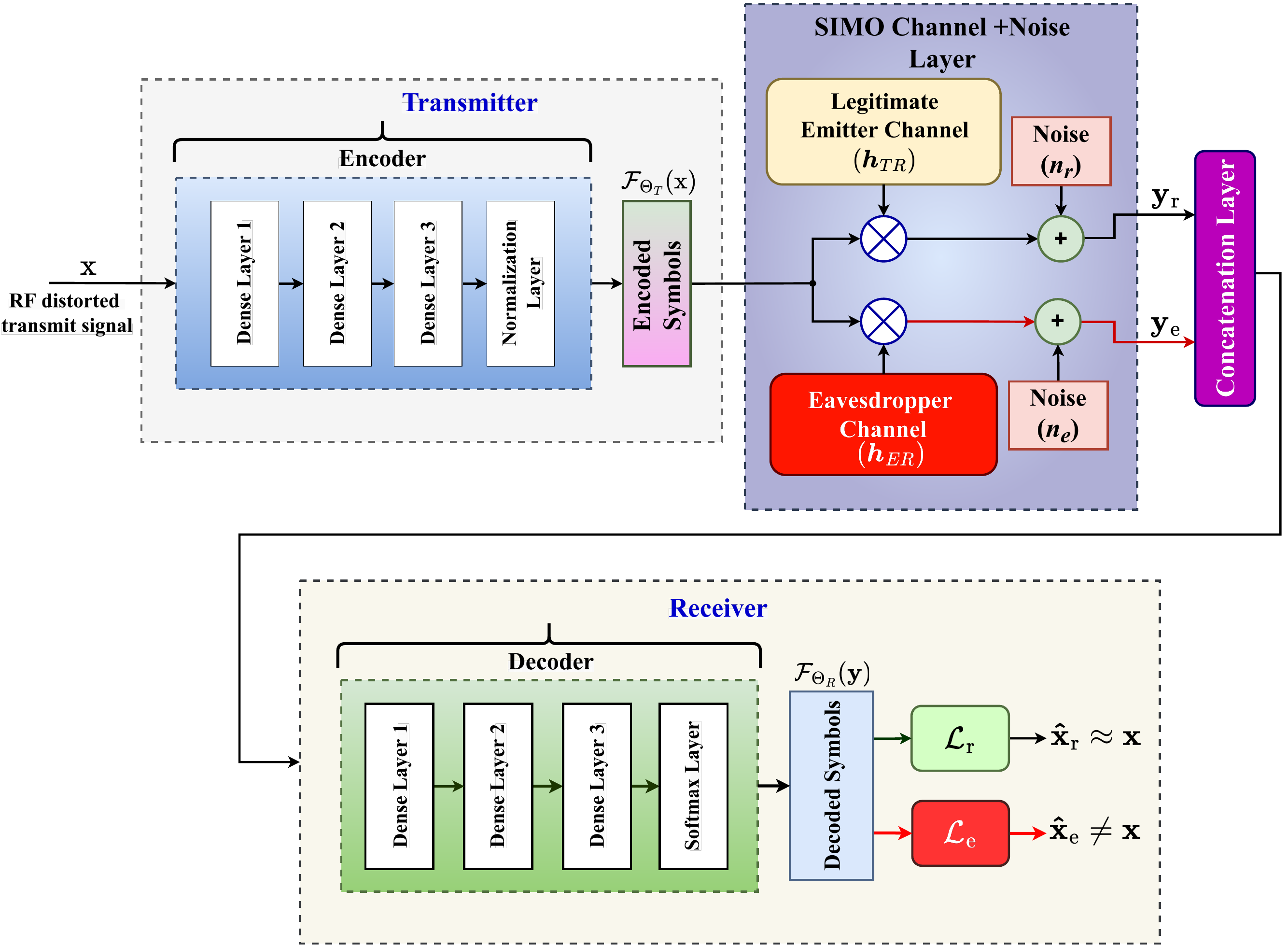}
    \caption{SIMO Auto-Encoder Model}
    \label{fig:SIMO_Autoenc}
\end{figure}

Our end-to-end communication system has one transmitter with a single antenna, a legitimate receiver, and an eavesdropper, both with multiple antennas and two wireless channels (the main and the wiretap channels). The transmitter aims to securely transmit a message ($\text{x}$) to the legitimate receiver via wireless channels while ensuring the eavesdropper cannot thwart the message. Despite both the legitimate receiver and the eavesdropper being able to receive the signal simultaneously, the transmission occurs through distinct wireless channels. The objective is to design a SIMO-based AE to facilitate secure end-to-end communication, with the aim of reconstructing the transmitted signal at the intended receiver with maximum probability $P_{r}\left ( \hat{\text{x}}_{r} = \text{x}|\textbf{y}_{r} \right )$. Concurrently, the design seeks to mimize the probability at the eavesdropper's $P_{e}\left ( \hat{\text{x}}_{e} = \text{x}|\textbf{y}_{e} \right ) $ end while ensuring comprehensive incorporation of all hardware impairments present at the RF front in the transmitted signal. The comprehensive architecture of the proposed AE is depicted in Fig. \ref{fig:SIMO_Autoenc}. A common loss function that could be used to reconstruct continuous input signals at the legitimate receiver is the Mean Squared Error (MSE). Conversely, the most suitable loss function for the eavesdropper should promote divergence between the eavesdropper's output and the originally transmitted signal. Theoretically, one appropriate loss function for this purpose is the negative MSE, aimed at maximizing the MSE between the decoded and transmitted signals. However, this is practically infeasible. Therefore, a softmax layer is employed as the AE output to address this issue, enabling cross-entropy as the loss function. This enables correct decoding by the legitimate receiver with the highest probability, expressed as.
\begin{equation}
    \mathcal{L}_{r}=-\sum_{i=1}^{K}\text{y}_{i}\log_{2}P_{i}
\end{equation}

In contrast, a legitimate receiver's optimization goal is to render it impossible for eavesdropping users to recover the original message accurately. Therefore, the optimal loss function should aim to equalize the decoder's output probabilities across the message space, thus maximizing the information entropy of the decoding result. Consequently, the loss function can be expressed as follows:
\begin{equation}
    \mathcal{L}_{e}=\sum_{i=1}^{K}{P}_{i}\log{P}_{i}
\end{equation}
The AE is trained using a joint loss function formulated for both the legitimate receiver and the eavesdropping user. This joint training combines the two components of the loss functions, enabling the optimization of the entire system's parameters through a unified sum-weighted loss function expressed as:
\begin{equation}
    \mathcal{L}_{total}=\alpha \cdot \mathcal{L}_{r}+(1-\alpha )\cdot\mathcal{L}_{e}
\end{equation}
The parameter $0\leq \alpha \leq 1$ balances the importance of minimizing the error at the legitimate receiver against maximizing the error at the eavesdropper, thus ensuring robustness against eavesdropping while maintaining reliable communication for the intended recipient. In this paper, we use $\alpha=0.5$ to ensure equal emphasis on both accurate decoding by the legitimate receiver and maximizing the confusion for the eavesdropper. However, while choosing higher values for legitimate receivers enhances their decoding ability, it will also lower the security against the eavesdropper.

\section{Experiment and Simulation Results}
The dataset comprises 50,000 samples of 16-QAM symbols sampled at a rate of 1 MHz. 70\% of the dataset is allocated for training, while the remaining 30\% is reserved for testing. The training is performed with a signal-to-noise ratio (SNR) ranging from 0 to 18 dB, drawn from a uniform distribution to enable the AE to learn across a wide range of SNR values. We adopt the FTR5123-B crystal LO \footnote{https://lora-alliance.org/sites/default/\%EF\%AC\%81les/showcase-documents/FTR5123-B0.pdf}, which is\ typically used in LoRa devices and mobile phones, with a frequency variation of $f_{ppm}=10\ \text{ppm}$. Table 1 summarizes the parameter values used in the simulation.
\begin{table}[!htb]  
\renewcommand{\arraystretch}{1.0}
\caption{Simulation settings}
\label{tab:experimental_parameter}
\centering
\begin{tabular}{lp{3.8cm}l}
    \hline
    Parameters & Values\\
    \hline
    Training samples &  35,000 \\
    \hline
    Test samples & 15,000 \\
    \hline
    Batch Size  & 256 \\
    \hline
    Transmit and Receive antennas  & $N_{t}=1$, $\bar{N}_{}=6$ \\
    \hline
    Gain and Phase imbalances & [−1 1] dB and [−5 5] degree\\
    \hline
    Sampling Frequency & 1 MHz\\
    \hline
    Carrier Frequency & 2 GHz\\
    \hline
    CFO & 1000Hz \cite{zhang2021radio}\\
    \hline
    $K_\text{vco}, V_\text{vco}$ & 100, 0.1 V\\
    \hline
    PA nonlinearity (Saleh model) & [$\alpha _{A}$=2.1587,$\beta _{A}$=1.1517, $\alpha _{P}$=4.0033$,\beta_{P}$=9.1040] \cite{zhang2021radio}\\
    \hline 
    Training SINR range & [0 - 18] dB \\
    \hline
    Test SINR range  & [0 - 22] dB \\
    \hline
    Initial Learning Rate  &  0.0003 \\
    \hline
    Learning Rate decay factor  &  0.65 \\
    \hline
    Training epoch &  100 \\
    \hline
\end{tabular}
\end{table}
\begin{figure}[ht]
    \centering
    \includegraphics[width=3.1in,height=2.3in]{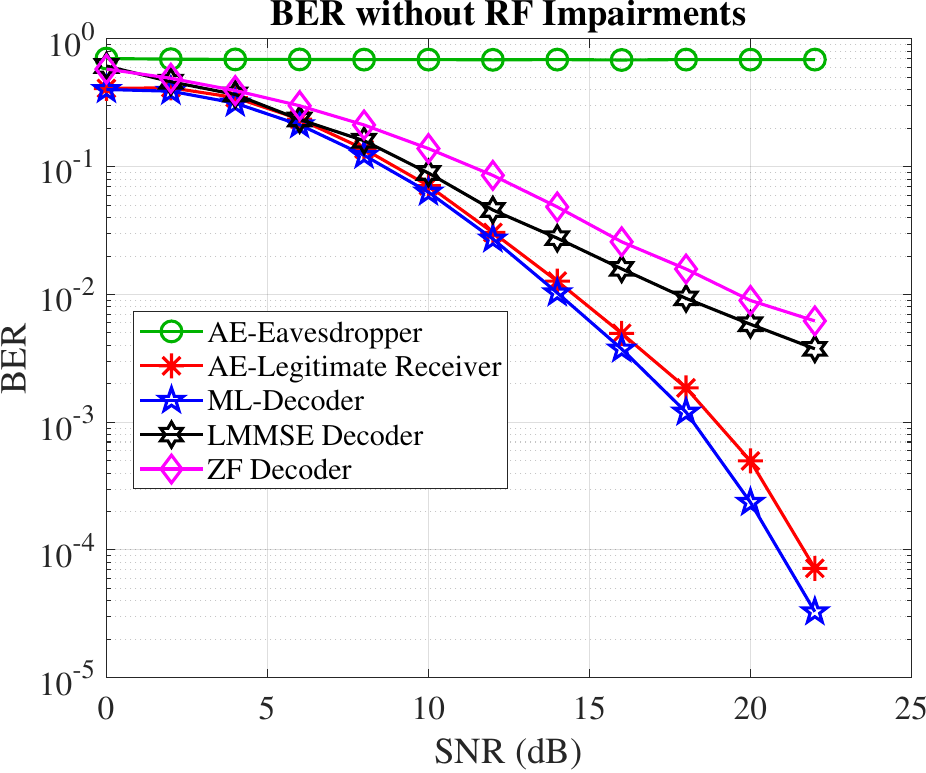}
    \caption{Symbol Error Rate (BER) without RF impairments}
    \label{fig:AE_Clean}
\end{figure}
\begin{figure}[ht]
    \centering
    \includegraphics[width=3.1in,height=2.3in]{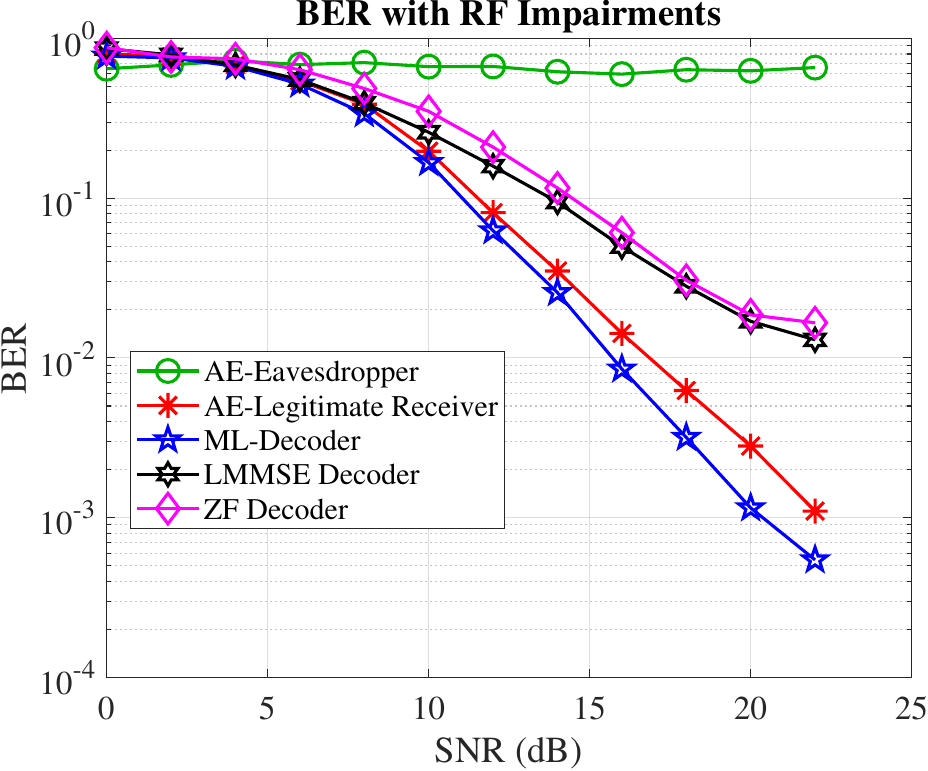}
    \caption{Symbol Error Rate (BER) with RF impairments}
    \label{fig:AE_Dist}
\end{figure}
 
Figs \ref{fig:AE_Clean} and \ref{fig:AE_Dist} illustrate the performance of classical linear decoders (ZR and LMMSE), the optimal nonlinear decoder (ML), and the proposed AE-based SIMO end-to-end learning model, both with and without RF hardware impairments. In the absence of impairments (Fig. \ref{fig:AE_Clean}), the AE-based model for the legitimate receiver shows the lowest BER across all SNR values near the optimal ML decoder with a minimal BER gap of less than $3.8969 \times 10^{-5}$ at SNR = 22 dB. Linear decoders like LMMSE and ZF exhibit higher BERs, highlighting their limitations.  Notably, the AE model for the eavesdropper The AE-consistently shows a high BER, indicating its inability to correctly decode the transmitted messages and thus ensuring secure communication. 

With RF impairments (Fig. \ref{fig:AE_Dist}), the AE-based model maintains its performance lead, although all decoders experience some degradation. The BER gap with the ML decoder at SNR = 22 dB increases to less than $5.5468 \times 10^{-4}$ below AE-based model. Linear decoders suffer more significantly, particularly ZF. The AE-eavesdropper's high BER persists, ensuring security. Additionally, the AE-based model ensures secure communication by maintaining a high BER for the eavesdropper, as observed by the constant line, thus preventing unauthorized decoding. Notably, the presence of RF impairments does not compromise this security feature; instead, it may enhance PHY security by making it even more challenging for eavesdroppers to intercept and decode the transmitted signals. 

\section{Conclusion}  
This paper introduces an AE-based SIMO model for end-to-end communication,  leveraging RF hardware impairments. A comprehensive mathematical signal model has been presented to encapsulate the hardware impairments inherent in various RF physical components. To assess the effectiveness of our proposed approach, we compared our model with traditional linear and optimal decoders in terms of SER performance. The experimental findings demonstrate that weighted joint autoencoder training ensures secure signal transmission, mitigating potential eavesdropping risks. In our future work, we will show that the drop in BER performance due to RF hardware impairments can be exploited to provide an additional layer of security, enabling the extraction of unique features present in the transmitted signal for enhanced PHY security and device identification.

\bibliographystyle{IEEEtran}
\bibliography{ref}
\end{document}